\PassOptionsToPackage{bookmarks={false}}{hyperref}
\documentclass[conference]{IEEEtran}
\IEEEoverridecommandlockouts
\usepackage[dvips]{graphicx}
\usepackage[fleqn]{amsmath}
\usepackage[varg]{txfonts}
\usepackage{colortbl}
\usepackage{algorithm}
\usepackage{algorithmic}
\usepackage{amsfonts}
\usepackage{tabularx}
\usepackage{cite}
\usepackage{graphicx} 
\ifCLASSINFOpdf
\else
\fi
 \PassOptionsToPackage{bookmarks={false}}{hyperref}
     \PassOptionsToPackage{draft}{hyperref}
\begin{document}

%
\title{Energy-Efficient Antenna Selection and Power Allocation for Large-Scale Multiple Antenna Systems with Hybrid Energy Supply}

\author{\IEEEauthorblockN{Zhenyu Zhou$^1$, Sheng Zhou$^2$, Jie Gong$^2$, and Zhisheng Niu$^2$}
\IEEEauthorblockA{$^1$State Key Laboratory of Alternate Electrical Power System with Renewable Energy Sources,\\
School of Electrical and Electronic Engineering, North China Electric Power University, Beijing, China, 102206\\
Email: zhenyu\_zhou@ncepu.edu.cn\\
$^2$Tsinghua National Laboratory for Information Science and Technology, Department of Electronic Engineering,\\
Tsinghua University, Beijing, China, 100084\\
Email: \{sheng.zhou, gongj13, niuzhs\}@tsinghua.edu.cn}
\thanks{This work is sponsored in part by the National Science Foundation of China (NSFC) under grant No. 61201191, the National Basic Research Program of China (973 Program: 2012CB316001), the Creative Research Groups of NSFC under grant No. 61321061, Fundamental Research Funds for the Central Universities under Grant No. 14MS08 and Hitachi, R\&D Headquarter.}}
\maketitle



%
\IEEEpeerreviewmaketitle

\begin{abstract}
The combination of energy harvesting and large-scale multiple antenna technologies provides a promising solution for improving the energy efficiency (EE) by exploiting renewable energy sources and reducing the transmission power per user and per antenna. However, the introduction of energy harvesting capabilities into large-scale multiple antenna systems poses many new challenges for energy-efficient system design due to the intermittent characteristics of renewable energy sources and limited battery capacity. Furthermore, the total manufacture cost and the sum power of a large number of radio frequency (RF) chains can not be ignored, and it would be impractical to use all the antennas for transmission.
In this paper, we propose an energy-efficient antenna selection and power allocation algorithm to maximize the EE subject to the constraint of user's quality of service (QoS). An iterative offline optimization algorithm is proposed to solve the non-convex EE optimization problem by exploiting the properties of nonlinear fractional programming. The relationships among maximum EE, selected antenna number, battery capacity, and EE-SE tradeoff are analyzed and verified through computer simulations.
\end{abstract}

\section{Introduction}

The information and communication technology (ICT) sector has been estimated to represent about 2 percent of the global $\mbox{CO}_2$ emissions \cite{global_footprint}, and 1.8 percent of the total world electricity consumption\cite{worldwide}. The mobile network operational expenditure (OPEX) for electricity globally is more than \$10 billion dollars, among which 80 percent of the energy is consumed at base stations (BSs) \cite{Dynamic_Niu}. As a result, energy-efficient communication technologies have received much attention in both industry and academic \cite{EE_Li}. Energy harvesting and large-scale multiple antennas are two emerging technologies for improving energy efficiency (EE). On one hand, energy harvesting that enables the BS to harvest energy from renewable energy sources such as solar, wind, and so on, can effectively reduce $\mbox{CO}_2$ emissions \cite{hybrid_energy_Gong}. On the other hand, large-scale multiple antenna systems which employ hundreds of antennas for transmission have been introduced to provide high spectral efficiency (SE) and reduce the transmission power per user and per antenna \cite{LMIMO_Marzetta, scaling_MIMO, LMIMO_Ngo}. Therefore, the combination of energy harvesting and large-scale multiple antenna technologies provides a promising solution for improving the EE by exploiting renewable energy sources and reduce the transmission power per user and per antenna. 

For energy harvesting systems, packet scheduling and power allocation algorithms have been proposed to minimize the transmission completion time, minimize the average grid power consumption or maximize the throughput (see \cite{Ozel_JSAC_2011, Yener2012, hybrid_energy_Gong} and references therein). However, most of these works target single-antenna systems, and few papers consider large-scale multiple antenna systems. A separate radio frequency (RF) chain is required for each employed antenna, which is usually more expensive than the antenna and does not follow Moore's law \cite{AS_MIMO}. As a result, the total manufacture cost and the sum power of a large number of RF chains can not be ignored, and it would be impractical and energy-inefficient to use all the antennas for transmission. In order to reduce the number of required RF chains, antenna selection techniques in multiple antenna systems have attracted intensive research interest (see \cite{AS_MIMO, antenna_Song, Min_BS} and references therein). However, these works are only valid for systems with a single energy source and are not applicable to the energy harvesting scenario.

The introduction of energy harvesting capabilities into large-scale multiple antenna systems poses many new challenges for energy-efficient system design due to the intermittent characteristics of renewable energy sources and limited battery capacity. In this paper, we propose an energy-efficient antenna selection and power allocation algorithm to maximize EE subject to the quality of service (QoS) constraint. We consider a more general hybrid energy supply model \cite{hybrid_energy_Gong, OFDM_EH_2013}, in which the BS is powered by both the conventional grid and renewable energy sources. The models which only consider the renewable energy \cite{ Ozel_JSAC_2011, Yener2012, hybrid_energy_Gong}, or the power grid \cite{antenna_Song, Min_BS, antenna_power_Liu}, can be regarded as special cases of the hybrid energy supply model. An iterative offline optimization algorithm is proposed to solve the non-convex EE optimization problem by exploiting the properties of nonlinear fractional programming \cite{Dinkelbach}. The relationships among maximum EE, selected antenna number, battery capacity, and EE-SE tradeoff are analyzed and verified through computer simulations.

The structure of this paper is organized as follows: Section \ref{System Model} introduces the system model and problem formulation in detail. Section \ref{offline} introduces the proposed energy-efficient antenna selection and power allocation algorithm. Section \ref{Simulation Results} introduces the simulation parameters, results and analyses. Section \ref{Conclusion} gives the conclusion and future works.

\section{System Model and Problem Formulation}
\label{System Model}

\subsection{System Model}
In the hybrid energy supply model, the harvested energy is first stored in a battery before it is used for data transmission. The power grid is required to compensate for the variability of the renewable energy sources to guarantee the QoS. We will not assume a particular type of renewable energy source in order to provide a general model for energy harvesting based communication systems. 

We adopt a similar system model as in \cite{OFDM_EH_2013, Ozel_JSAC_2011} by modeling the channel fading and energy harvesting as stochastic processes. The energy arrival times in the BS are modeled as a Poisson counting process with rate $\lambda_E$, and the block fading channel model is assumed. Different from \cite{OFDM_EH_2013, Ozel_JSAC_2011}, we assume that the energy harvesting rate changes slowly (several seconds) compared to the communication block length (several milliseconds) \cite{EH_Huang_2014, Huang_INFOCOM_2014}. Therefore, the energy harvesting rate $\lambda_E$ could be treated as identical over thousands of communication blocks.

 The energy arrivals occur in countable time instants, which are indexed as $\{t_1^E, t_2^E, \ldots \}$, and the inter-occurrence time between any two consecutive energy arrival events, i.e., $t_i^E-t_{i-1}^E, i \in \{1, 2, \ldots\}$, is exponentially distributed with mean $1/\lambda_E$ by the Poisson property. We assume that $E_0$ units of energy are available at time $t_0^E=0$. For energy arrival events happened at time instants $\{t_1^E, t_2^E, \ldots\}$, $\{E_1, E_2, \ldots\}$ units of energy are harvested respectively. We will refer to the time interval between two consecutive energy arrival events as an ``epoch". For a total duration of $[0, T_{total}]$, if $L$ energy arrival events happened, there is a total of $L$ epochs for the considered duration of $T_{total}$ seconds. Epoch $i$, $i \in \{1, 2, \ldots, L\}$, is defined as the time interval $[t_{i-1}^E, t_i^E)$, and the length of the epoch $i$ is defined as $T_i= t_i^E-t_{i-1}^E$. The energy harvested in epoch $i$ is defined as $E_{in}[i]$.

We consider a  typical downlink cellular system, in which data are transmitted from the BS to mobile terminals. The BS is equipped with a total of $N$ $(N>>1)$ antennas and the mobile terminal has only one antenna, which is common in the real world. The case of multiple antennas at the receiver will be discussed in future works. The received signal at the mobile terminal can be written as
\begin{equation}
y=\mathbf{H}^T\mathbf{X}+n,
\end{equation}
where $\mathbf{X}$ represents the $N$-dimensional precoded transmitted symbol, i.e., $\mathbf{X}=\frac{\mathbf{H}^{*}}{\parallel \mathbf{H}  \parallel }x$, $n$ is the additive Gaussian white noise (AWGN) with the mean zero and variance $\sigma^2$ normalized to $1$. $\mathbf{H}=[h_1,h_2,\ldots,h_N]^T$ is the $N \times 1$ vector of channel gains with the element $h_j$ representing the gain from the transmit antenna $j$ to the mobile terminal.

In order to reduce the number of RF chains, an energy-efficient transmitter antenna selection algorithm is required to choose the best $M$ $(1 \leq M \leq N)$ antennas from all the available $N$ antennas. We assume that perfect channel state information (CSI) is known at the transmitter. How to obtain CSI is out of the scope of this paper and is not considered here. According to (5.31) in \cite{Wireless_Tse}, the achievable rate of transmit antenna selection $I_{sel}$ (bits/s/Hz) is given by
\begin{equation}
I_{sel}=\mbox{log}_2  (1+P_{Tx}\sum_{j=1}^M|h_j|^2),
\end{equation} 
where $|h_1|^2>|h_2|^2>...>|h_M|^2$, and $P_{Tx}$ is the total power constraint across the transmission antennas. Due to the channel hardening phenomenon in antenna selection systems \cite{antenna_Song}, the mutual information for large $N$ and $1 \leq M \leq N$ has a folded normal distribution, which is given by
\begin{equation}
 \label{eq:I_sel}
I_{sel} \sim \mathcal{FN} \bigg( \mbox{log}_2\left[1+\big( 1+\mbox{In}\frac{N}{M}\big)P_{Tx}M \right], \frac{(\mbox{log}_2e P_{Tx})^2 M (2-\frac{M}{N})}{(1+(1+\mbox{In}\frac{N}{M})P_{Tx}M)^2} \bigg).
\end{equation}
Although (\ref{eq:I_sel}) is derived with the assumption that $N$ and $M$ are large, simulation results in \cite{antenna_Song} demonstrate that it also works well when $N$ and $M$ are ``not so large".  

\subsection{Problem Formulation}
\label{Problem Formulation}
In this subsection, we consider the weighted EE over a total of $L$ epochs, $U_{EE}$ (bits/Hz/Joule), which is defined as 
\begin{equation}
 \label{eq:U_EE}
U_{EE}=\frac{U_{SE}}{E^W_{total}}. 
\end{equation}
The total spectral efficiency, $U_{SE}$ (bits/Hz), is given by
\begin{align}
\label{eq:U_SE}
U_{SE}&=\sum_{i=1}^L \mathbb{E}\bigg[ I_{sel}[i]   \bigg] T_i\notag\\
&=\sum_{i=1}^L \left(  \mbox{log}_2\left[1+\bigg( 1+\mbox{In}\frac{N}{M[i]}\bigg)P_{Tx}[i]M[i] \right]  \right) T_i,
\end{align}
where $\mathbb{E}\bigg[ I_{sel}[i]   \bigg]$ denotes the expectation of the mutual information. The weighted total energy consumption of the BS, $E_{total}^W$ (Joule), is given by
\begin{align}
 \label{eq:E_total}
E_{total}^W&=
&=\sum_{i=1}^L \left(
P_C^W[i]+\frac{1}{\eta} P_{Tx}^W[i]+M[i]P_{RF}^W[i] \right) T_i,
\end{align}
 where $P_C^W$ is the weighted constant circuit power,  $P_{Tx}^W$ is the weighted transmission power, $P_{RF}^W$ is the weighted RF chain power consumption which includes mixer, active filters, digital to analog converter (DAC), etc, and $\eta$ is the power amplifier (PA) efficiency, i.e., $0<\eta<1$. In the considered hybrid energy supply model, the BS is powered by both the renewable energy and the power grid. Therefore, $P_C^W$,  $P_{Tx}^W$, and $P_{RF}^W$ can be modeled as 
 \begin{align}
   \label{eq:P_C}
 P_C^W[i]&=wP_C^E[i]+P_C^G[i],\\
 \label{eq:P_t}
  P_{Tx}^W[i]&=wP_{Tx}^E[i]+P_{Tx}^G[i],\\
  \label{eq:P_RF}
  P_{RF}^W[i]&=wP_{RF}^E[i]+P_{RF}^G[i],
 \end{align}
where $P_C^E$ and $P_C^G$ are the instantaneous circuit power drawn from the renewable source and the power grid respectively, $P_{Tx}^E$ and $P_{Tx}^G$ are the instantaneous transmission power drawn from the renewable source and the power grid respectively, $P_{RF}^E$ and $P_{RF}^G$ are the instantaneous RF chain power drawn from the renewable source and the power grid respectively. $w$ reflects either a normalized physical cost or a normalized virtual cost with regards to the usage of the power grid \cite{OFDM_EH_2013}. In this paper, $w$ is set as $0<w<1$ to encourage the BS to consume more renewable energy. 

The set of antenna selection solutions is defined as $\mathcal{S}=\{M[i], \forall i \in [1,L]\}$, and the set of power allocation solutions is defined as $\mathcal{P}=\{P_C^E[i], P_C^G[i], P_{Tx}^E[i], P_{Tx}^G[i], P_{RF}^E[i], P_{RF}^G[i], \forall i \in [i,L] \}$. Taking (\ref{eq:U_SE}), (\ref{eq:E_total}), (\ref{eq:P_C}), (\ref{eq:P_t}), (\ref{eq:P_RF}) into (\ref{eq:U_EE}), the weighted EE is given as
\begin{align}
\label{eq:U_EE_full}
U_{EE}(\mathcal{S}, \mathcal{P})=\frac{U_{SE}(\mathcal{S}, \mathcal{P})}{E^W_{total} (\mathcal{S}, \mathcal{P})},
\end{align}
where
 \begin{align}
\label{eq:U_SE_full}
&U_{SE}(\mathcal{S}, \mathcal{P})=\notag\\
&\sum_{i=1}^L \mbox{log}_2\bigg[1+\big( 1+\mbox{In}\frac{N}{M[i]}\big)\big( P_{Tx}^E[i]+P_{Tx}^G[i]   \big) M[i] \bigg]T_i,
\end{align}
\begin{align}
\label{eq:E_total_full}
&E^W_{total} (\mathcal{S}, \mathcal{P})=\notag\\
&\sum_{i=1}^L \Bigg( w\bigg(P_C^E[i]+\frac{1}{\eta}P_{Tx}^E[i]+M[i]P_{RF}^E[i]\bigg)+P_C^G[i]\notag\\
&+\frac{1}{\eta}P_{Tx}^G[i]+M[i]P_{RF}^G[i] \Bigg) T_i.
\end{align}

The EE optimization problem can be formulated as
\begin{align}
 \label{eq:optimization problem}
&\max_{(\mathcal{S}, \mathcal{P})}. \hspace{10mm} U_{EE}\left( \mathcal{S}, \mathcal{P} \right)\notag\\
&\mbox{s.t.} \hspace{10mm} C1, C2, C3, C4, C5, C6, C7, C8, C9.
\end{align}
\begin{align}
C1: &\sum_{i=1}^{e} \left( \frac{1}{\eta}P_{Tx}^E[i]+P_C^E[i]+M[i]P_{RF}^E[i]\right) T_i \notag\\
&\leq \sum_{i=1}^{e}E_{in}[i], \forall e,
\end{align}
\begin{align}
C2:&\sum_{i=1}^{e}E_{in}[i]-\sum_{i=1}^{e-1}\bigg( \frac{1}{\eta}P_{Tx}^E[i]+P_C^E[i]+M[i]P_{RF}^E[i]\bigg) T_i \notag\\ 
&\leq B_{max}, \forall e,
\end{align}
\begin{align}
C3:(P_C^E[i]+P_C^G[i])T_i=P_CT_i, \forall i,
\end{align}
\begin{align}
C4: (P_{RF}^E[i]+P_{RF}^G[i])M[i]T_i=P_{RF}M[i]T_i, \forall i,
\end{align}
\begin{align}
C5:(P_{Tx}^E[i]+P_{Tx}^G[i])T_i \leq P_{Tx,max}T_i, \forall i, 
\end{align}
\begin{align}
C6:\left( \frac{1}{\eta}P_{Tx}^G[i]+P_C^G[i]+M[i]P_{RF}^G[i] \right) T_i \leq P_{max}^GT_i, \forall i,
\end{align}
\begin{align}
\label{eq:QOS}
&C7:\sum_{i=1}^{L}\mathbb{E}\bigg[I_{sel}[i]\bigg] T_i \geq R_{min},\\
&C8:1\leq M \leq N,\\
&C9:P_C^E[i], P_C^G[i], P_{Tx}^E[i], P_{Tx}^G[i], P_{RF}^E[i], P_{RF}^G[i] \geq 0, \forall i.
\end{align}
The constraint C1 specifies the causality constraint, i.e., energy that has not been harvested yet cannot be used at the current time. C2 specifies the battery capacity constraint in order to prevent energy overflow. C3 ensures that the energy required for BS circuit operation is always available. C4 ensures that the energy required for a total number of $M$ RF chains is always available. C5, C6 are constraints on the maximum transmission power of the BS and the maximum supplying power of the grid respectively. C7 specifies the QoS requirement in terms of minimum transmission rate. C8 is the antenna selection range constraint and C9 is the non-negative constraint on the power allocation variables.

\section{The Energy-efficient Antenna Selection and Power Allocation Algorithm}
\label{offline}

\subsection{The Objective Function Transformation}
\label{transformation}

The optimization problem in (\ref{eq:optimization problem}) is non-convex due to the fractional form. We transformed the fractional objective function to a subtractive function by using the nonlinear fractional programming developed in \cite{Dinkelbach}. We define the maximum weighted EE as $q^{*}$, which is given by
\begin{equation}
q^{*}=\max. U_{EE}{(\mathcal{S},\mathcal{P})} =\frac{U_{SE}(\mathcal{S}^{*},\mathcal{P}^{*})}{E_{total}^W(\mathcal{S}^{*},\mathcal{P}^{*})},
\end{equation} 
where $(\mathcal{S}^{*},\mathcal{P}^{*})$ is the optimum antenna selection and power allocation policy. The following theorem can be proved:

\textbf{\emph{Theorem 1:}} The maximum weighted EE $q^{*}$ is achieved if and only if 
\begin{align}
&\max. \:\:U_{SE}(\mathcal{S},\mathcal{P})-q^{*}E_{total}^W(\mathcal{S},\mathcal{P})\nonumber\\
&=U_{SE}(\mathcal{S}^{*},\mathcal{P}^{*})-q^{*}E_{total}^W(\mathcal{S}^{*},\mathcal{P}^{*})=0.
\end{align}

\begin{IEEEproof}
The proof of Theorem 1 is similar to the proof of the Theorem (page 494 in \cite{Dinkelbach}).
\end{IEEEproof}

\textbf{\emph{Corollary 1:}} For each fixed $\mathcal{P}$, the transformed objective function in subtractive form, i.e.,$U_{SE}(\mathcal{S},\mathcal{P})-qE_{total}^W(\mathcal{S},\mathcal{P})$, is a concave function with regards to $\mathcal{S}$. For each fixed $\mathcal{S}$, the transformed objective function in subtractive form is jointly concave with regards to all the optimization variables in $\mathcal{P}$.

\begin{IEEEproof}
The proof of Corollary 1 is given in Appendix \ref{corollary1}.
\end{IEEEproof}


\subsection{The Iterative Offline Optimization Algorithm}
\label{algorithm}

The proposed algorithm is summarized in Algorithm \ref{offline algorithm}. $n$ is the iteration index, $L_{max}$ is the maximum number of iterations, and $\Delta$ is the maximum tolerance. 
At each iteration, for any given $q$, the corresponding resource allocation solution $(\mathcal{S},\mathcal{P})$ is obtained by solving the following transformed optimization problem:
\begin{align}
 \label{eq:transformed problem}
 &\max . \:\: U_{SE}(\mathcal{S},\mathcal{P}) -qE_{total}^W(\mathcal{S},\mathcal{P})\nonumber\\
 &\mbox{s.t.} \:\:\: C1, C2, C3, C4, C5, C6, C7, C8, C9.
\end{align}
The Lagrangian associated with the problem (\ref{eq:transformed problem}) is given by
\begin{align}
&\mathcal{L}(\mathcal{S},\mathcal{P}, \alpha, \beta, \gamma, \delta, \zeta, \theta, \mu)=U_{SE}(\mathcal{S},\mathcal{P}) -qE_{total}^W(\mathcal{S},\mathcal{P})\notag\\
&-\sum_{i=1}^L \alpha_i  \bigg( \sum_{k=1}^i \big( \frac{1}{\eta}P_{Tx}^E[k]+P_C^E[k]+M[k]P_{RF}^E[k] \big) T_k- \sum_{k=1}^i E_{in}[k] \bigg)\notag\\
&-\sum_{i=2}^{L+1}\beta_i \bigg(  \sum_{k=1}^i E_{in}[k] -\sum_{k=1}^{i-1}\big( \frac{1}{\eta}P_{Tx}^E[k]+P_C^E[k]+M[k]P_{RF}^E[k]\big) T_k\notag\\
&-B_{max} \bigg)+\sum_{i=1}^L \gamma_i \left(P_C^E[i]+P_C^G[i]-P_C \right) T_i+\sum_{i=1}^L \delta_i \bigg( P_{RF}^E[i]\notag\\
&+P_{RF}^G[i])M[i]-P_{RF}M[i] \bigg) T_i-\sum_{i=1}^L \zeta_i \bigg( P_{Tx}^E[i]+P_{Tx}^G[i]\notag\\
&-P_{Tx,max} \bigg) T_i-\sum_{i=1}^L \theta_i \bigg( \frac{1}{\eta}P_{Tx}^G[i]+P_C^G[i]+M[i]P_{RF}^G[i]- P_{max}^G\bigg) T_i  \notag\\
&+\mu \left( U_{SE}(\mathcal{S},\mathcal{P})-R_{min} \right),
\end{align}
where $\alpha, \beta, \gamma, \delta, \zeta, \theta, \mu$ are the Lagrange multipliers associated with constraints C1-C7 respectively. The equivalent dual problem can be decomposed into two parts: the maximization problem solves the resource allocation problem and the minimization problem solves corresponding Lagrange multipliers, which is given by
\begin{equation}
\label{eq:dual problem}
 \displaystyle \min_{\displaystyle (\alpha, \beta, \gamma, \delta, \zeta, \theta, \mu \geq 0)}  \max_{\displaystyle (\mathcal{S},\mathcal{P})} \mathcal{L}(\mathcal{S},\mathcal{P}, \alpha, \beta, \gamma, \delta, \zeta, \theta, \mu) 
\end{equation}
From Corollary 1, we know that the objective function in (\ref{eq:transformed problem}) is concave over $\mathcal{P}$ with $\mathcal{M}$ fixed. The Karush-Kuhn-Tucker (KKT) conditions are used to find the optimum power allocation solutions. For any given $q$, the corresponding optimum solution is given by
\begin{equation}
\label{eq:P_TE_optimum}
\hat{P}_{Tx}^{E}[i] =\left[ \frac{(1+\mu )\eta  \mbox{log}_2e}{qw+\sum_{k=i}^{L}\alpha_k -\sum_{k=i}^{L}\beta_{k+1} +\eta \zeta_i }-\Phi[i]  \right]^{+},
\end{equation}
\begin{equation}
\label{eq:P_TG_optimum}
\hat{P}_{Tx}^{G}[i] = \left[\frac{(1+\mu )\eta  \mbox{log}_2e}{q+\eta \zeta_i +\theta_i  }-\Phi[i]-P_t^{E}[i] \right]^{+},
\end{equation}
\begin{equation}
\label{eq:P_CE_optimum}
\hat{P}_C^{E}[i] =\left[ \frac{E_{C}[i]}{T_i}\right]_0^{P_C},
\end{equation}
\begin{equation}
\label{eq:P_CG_optimum}
\hat{P}_C^{G}[i] =P_C-\hat{P}_C^{E}[i],
\end{equation}
\begin{equation}
\label{eq:P_RFE_optimum}
\hat{P}_{RF}^{E}[i] =\left[ \frac{E_{RF}[i]}{\hat{M}[i]T_i}\right]_0^{P_{RF}},
\end{equation}
\begin{align}
\label{eq:P_RFG_optimum}
\hat{P}_{RF}^{G}[i] =P_{RF}-\hat{P}_{RF}^{E}[i],
\end{align}
where
\begin{align}
\Phi[i] &=\frac{1}{(1+\mbox{In}\frac{N}{\hat{M}[i]})\hat{M}[i]},\\
E_{C}[i]&=\sum_{k=1}^{i}E_{in}[k]-\sum_{k=1}^{i}\frac{1}{\eta}\hat{P}_{Tx}^{E}[k]T_k-\sum_{k=1}^{i-1}\hat{M}[k]\hat{P}_{RF}^{E}[k]T_k,\notag\\
&-\sum_{k=1}^{i-1} \hat{P}_{C}^E[k]T_k\\
E_{RF}[i]&=\sum_{k=1}^{i}E_{in}[k]-\sum_{k=1}^{i}(\frac{1}{\eta}\hat{P}_{Tx}^{E}[k]+\hat{P}_C^E[k])T_k\notag\\
&-\sum_{k=1}^{i-1}\hat{M}[k]\hat{P}_{RF}^{E}[k]T_k.
\end{align}
$[x]^+=\max\{0,x\}$. $[x]_b^a=a$, if $x>a$; $[x]_b^a=x$, if $b \leq x \leq a$; $[x]_b^a=b$, if $x<b$. $E_C[i]$ and $E_{RF}[i]$ represents the residual energy level in the battery. (\ref{eq:P_TE_optimum}), (\ref{eq:P_TG_optimum}) indicates a water-filling algorithm for transmission power allocation, and $P_{Tx}^{E}$ decreases the water level of $P_{Tx}^{G}$ by reducing the amount of energy drawn from the power grid. (\ref{eq:P_CE_optimum}), (\ref{eq:P_CG_optimum}) indicates that if the residual energy in the battery is not sufficient to support the required circuit energy $P_CT_i$, i.e., $P_C^{E}[i]<P_C$, then the BS will draw $P_C^{G}[i]T_i$ energy from the power grid. Similar analysis can be obtained from (\ref{eq:P_RFE_optimum}) and (\ref{eq:P_RFG_optimum}) for the circuit power allocation of RF chains. 
By solving the optimization problem for a given $\mathcal{P}$, we can obtain the maximum objection value for each combination of feasible $(\mathcal{\hat{S}}, \mathcal{P})$, and then choose the pair with the maximum value among all possible combinations. The optimum $\mathcal{\hat{P}}$ can be obtained by bisection method \cite{convex_optimization}.

\begin{algorithm}[t]
\caption{Iterative Offline Optimization Algorithm}
\label{offline algorithm}
\begin{algorithmic}[1]
\STATE $q \leftarrow 0$, $L_{max} \leftarrow 10$, $n \leftarrow 1$, $\Delta \leftarrow 10^{-3}$ 
\FOR{$n=1$ to $L_{max}$}
\STATE solve the optimization problem in (\ref{eq:transformed problem}) for a given $q$ and obtain $(\mathcal{\hat{S}},\mathcal{\hat{P}})$
\IF{$U_{SE}(\mathcal{\hat{S}},\mathcal{\hat{P}})-qE_{total}^W(\mathcal{\hat{S}},\mathcal{\hat{P}}) \leq \Delta $,}
\STATE $(\mathcal{S}^*,\mathcal{P}^*)=(\mathcal{\hat{S}},\mathcal{\hat{P}})$, and $\displaystyle q^{*}=\frac{U_{SE}(\mathcal{S}^*,\mathcal{P}^*)}{E_{total}^W(\mathcal{S}^*,\mathcal{P}^*)}$
\STATE \textbf{break}
\ELSE
\STATE $\displaystyle  q=\frac{U_{SE}(\mathcal{\hat{S}},\mathcal{\hat{P}})}{E_{total}^W(\mathcal{\hat{S}},\mathcal{\hat{P}})}$, and $n=n+1$
\ENDIF
\ENDFOR
\end{algorithmic}
\end{algorithm}

For solving the minimization problem, the Lagrange multipliers can be updated by using the gradient method \cite{improved_step_size}. More details about the Lagrange multipliers updating, complexity analysis, convergence analysis, and implementation are described in future journal version.

\begin{table}[t]
\caption{Simulation Parameters.}
\label{simulation_parameters}
\begin{center}
\begin{tabular}{|l|l|}
\hline
\textbf{Parameter}&\textbf{Value}\\
\hline
Maximum transmission power $P_{Tx,max}$ & 46 dBm\\
\hline
Virtual cost of renewable energies $w$ & 0.01\\
\hline
Constant circuit power $P_C$ &160.8 W\\
\hline
RF chain circuit power $P_{RF}$ & 160 mW \\
\hline
Total number of antennas $N$ & 100\\
\hline
Maximum grid power $P_{max}^G$ & 300 W\\
\hline
PA efficiency $\eta $ & 35\%\\
\hline
Duration $T_{total}$ & 7 s\\
\hline
QoS $R_{min}$ & 7 bits/Hz\\
\hline
\end{tabular}
\end{center}
\end{table}

\section{Simulation Results}
\label{Simulation Results}

In this section, the proposed algorithm is verified through computer simulations. The simulation parameter values are inspired by \cite{Earth, AS_Cimini}, and are summarized in Table \ref{simulation_parameters}. It is noted that the weight $w$ does not affect the optimal antenna selection and power allocation solution as long as $0<w<1$. However, the weighted EE is indeed affected by $w$. Hence, $w$ is fixed as $0.01$ throughout the simulations for the purpose of fair comparison.   

\begin{figure}[t]
\begin{center}
\scalebox{0.55} 
{\includegraphics{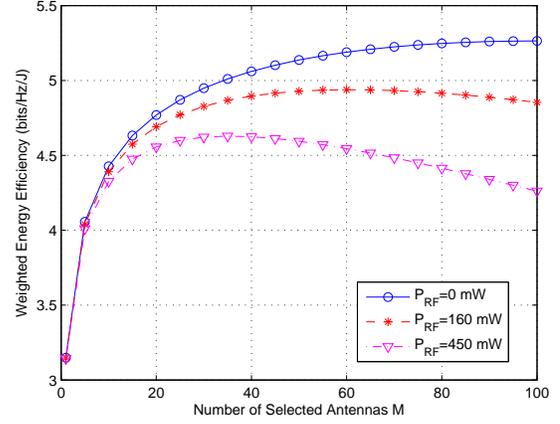}}
\end{center}
\caption{The weighted energy efficiency $U_{EE}$ corresponding to the number of selected antennas $M$  ($B_{max}=1500$ J, $E_{in}=1000$ J $P_{RF}=0, 160, 450$ mW, $T_{total}=3$ s).}
\label{EE_M}
\end{figure}

Fig. \ref{EE_M} shows the weighted EE $U_{EE}$ corresponding to the number of selected antennas $M$ with different RF chain circuit power $P_{RF}$. The values of the harvested energy and battery capacity are just taken for illustration purpose, i.e., $E_{in}=1000$ J and $B_{max}=1500$ J. Each curve is simulated by using a different $P_{RF}$, with $P_{RF}=0$ mW represents the ideal RF chain that is energy free. For the case of $P_{RF}=0$ mW, $U_{EE}$ increases monotonically with $M$. However, for the case of $P_{RF}=160$ mW and  $P_{RF}=450$ mW, $U_{EE}$ increases first and then decreases as $M$ increases, and the optimum number $M^*$ is 61 and 35 respectively. It is not energy efficient to use all of the available antennas for transmission. Besides, both the optimal EE $U_{EE}^*$ and selected antenna number $M^*$ decreases as the RF chain circuit power $P_{RF}$ increases.

\begin{figure}[t]
\begin{center}
\scalebox{0.55} 
{\includegraphics{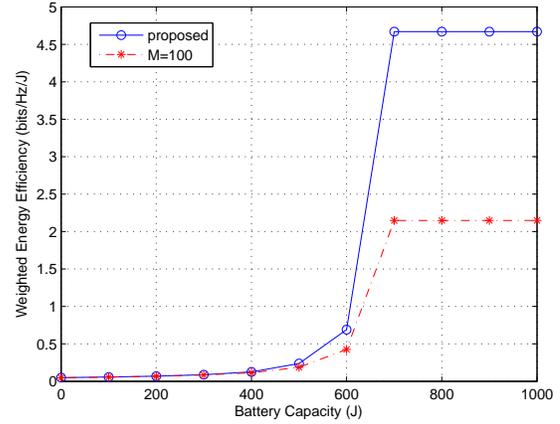}}
\end{center}
\caption{The relationship between the battery capacity and the weighted energy efficiency ($B_{max}=1500$ J, $E_{in}=700$ J, $P_{RF}=160$ mW, $T_{total}=7$ s).}
\label{EE_capacity}
\end{figure}

\begin{figure}[t]
\begin{center}
\scalebox{0.55} 
{\includegraphics{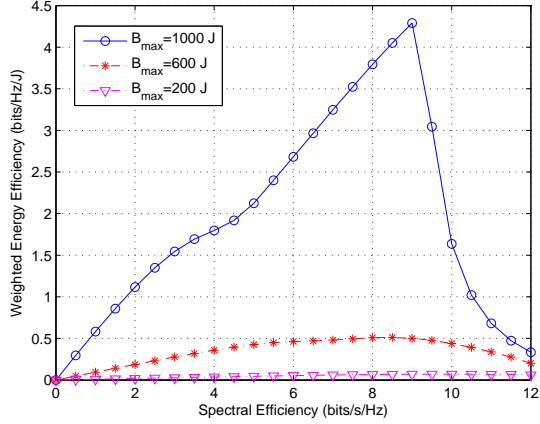}}
\end{center}
\caption{The energy efficiency and spectral efficiency tradeoff under different battery capacity conditions. ($B_{max}=1000, 600$ J, $E_{in}=700$ J, $P_{RF}=160$ mW, $T_{total}=7$ s).}
\label{EE_SE_tradeoff}
\end{figure}

The impact of battery capacity $B_{max}$ on the weighted EE is investigated in Fig. \ref{EE_capacity}. $B_{max}$ is increased from $0$ J to $1000$ J with a step of $100$ J, and for each $B_{max}$, the corresponding optimum weighted EE $U_{EE}^*$ is obtained by Algorithm \ref{offline algorithm}. The energy overflow constraint C2 is removed. The proposed algorithm (labeled as ``proposed") is compared with the strategy that uses all of the available antennas for transmission (labeled as ``$M=100$"). We can see that $U_{EE}^*$ increases monotonically with $B_{max}$, until to the condition that the system is no longer limited by the battery capacity. The proposed algorithm significantly outperforms the algorithm with $M=100$ and can improve the EE by more than $110\%$. The reason is further explained in Fig. \ref{EE_SE_tradeoff}. For the case that $B_{max} \leq 600$ J, the improvement brought by the proposed algorithm is not obvious due to the fact that the maximum achievable EE is limited by the battery capacity.   

Fig. \ref{EE_SE_tradeoff} shows the tradeoff between EE and SE under three different battery capacity conditions, i.e., $B_{max}=1000, 600, 200$ J respectively. SE is increased from $0$ bits/s/Hz to $12$ bits/s/Hz with a step of $0.5$, and the corresponding EE is obtained through computer simulations. The inequality QoS constraint defined in (\ref{eq:QOS}) is reduced to an equality constraint subject to the given SE. For the case of $B_{max}=1000$ J, the maximum achievable SE and EE subject to the constraints (defined in problem  (\ref{eq:optimization problem}))  are $12$ bits/s/Hz and $4.2898$ bits/s/J respectively. In comparison, for the case of $B_{max}=600$ J, the maximum achievable SE and EE are $11$ bits/s/Hz and $0.5113$ bits/s/J respectively. By decreasing the battery capacity from $1000$ J to $600$ J, the maximum achievable SE and EE are reduced by nearly  $8\%$ and $88\%$ respectively. It is clear that the limited battery capacity has a much more severe impact on EE than on SE due to energy overflow. In particular, for the case of $B_{max}=1000$ J, if we increase the SE from $9$ bits/s/Hz to $12$ bits/s/Hz ( $30\%$ improvement), the corresponding EE is reduced by more than $92\%$. Hence, increasing transmission power beyond the power for optimal EE brings little SE improvement but significant EE loss. Similar observations have also been found in \cite{EE_SE_OFDM2014} with considering practical power amplifier saturation. However, in the battery limited case, the EE loss is not so large due to the fact that the maximum achievable EE is limited by the battery capacity.   

\section{Conclusion and Future Works}
\label{Conclusion}

In this paper, an iterative offline antenna selection and power allocation algorithm was proposed for large-scale multiple antenna systems with hybrid energy supply. The relationships among energy efficiency, selected antenna number, battery capacity, and EE-SE tradeoff were analyzed and verified through computer simulations. In practice, since the future energy arrival information is not available, dynamic programming (DP) based optimal online optimization policy should be studied. However, due to the ``curse of dimensionality" associated with DP, future works should be focused on suboptimal algorithms with low computation complexity and close-to-optimal performance.

\appendices

\section{Proof of the Corollary 1}
\label{corollary1}

Firstly, let us prove the first part of Corollary 1. For each fixed $\mathcal{P}$, we consider the transformed function as a function $\mathcal{S}$. Taking the second-order derivative of $U_{SE}$ (defined in \ref{eq:U_SE_full}) with regards to $M[i]$, the denominator of $ \frac{\partial^2 U_{SE}}{\partial (M[i])^2}$ is surely a positive value, and the numerator $G[i]$ is 
\begin{align}
G[i]=-\frac{\mbox{log}_2 e }{M[i]}P_{Tx}[i]T_i\Gamma [i]-\left(\mbox{In}\frac{N}{M[i]}P_{Tx}[i]  \right)^2 T_i \mbox{log}_2 e < 0,
\end{align}
where $\Gamma [i]=\left( 1+\big( 1+\mbox{In}\frac{N}{M[i]}\big)P_{Tx}[i]M[i]  \right)>0$. Thus, we have $ \frac{\partial^2 U_{SE}}{\partial (M[i])^2} <0, \forall i$, and proves that $U_{SE}$ is a concave function of $M$. Similarly, it can be easily proved that $-qE_{total}^W$ is an affine function of $M$. Since the sum of a concave function and an affine function is also concave, this completes the proof of the first part of Corollary 1.

Secondly, for each fixed $M[i]$, we consider the transformed function as a function $\mathcal{P}$. Since $U_{SE}$ is a logarithmic function of $P_{Tx}^E$ and $P_{Tx}^G$, $U_{SE}$ is jointly concave with $P_{Tx}^E$ and $P_{Tx}^G$ \cite{convex_optimization}. On the other hand, $-qE_{total}^W$ is an affine function of $P_{Tx}^E, P_{Tx}^G, P_{C}^E, P_{C}^G, P_{RF}^E, P_{RF}^G$.  Since the sum of a concave function and an affine function is also concave, this completes the proof of the second part of Corollary 1.

\bibliographystyle{IEEEtran}
\bibliography{IEEE_gc_2014}


\end{document}